# Schooling to Exploit Foolish Contracts

Tamer Abdelaziz
*National University of Singapore*
Singapore
tamer@comp.nus.edu.sg

Aquinas Hobor
*University College London*
London, England
a.hobor@ucl.ac.uk

*Abstract*—We introduce SCooLS, our <u>S</u>mart <u>C</u>ontract Learning (<u>S</u>emi-supervised) engine. SCooLS uses neural networks to analyze Ethereum contract bytecode and identifies specific vulnerable functions. SCooLS incorporates two key elements: semi-supervised learning and graph neural networks (GNNs). Semi-supervised learning produces more accurate models than unsupervised learning, while not requiring the large oracle-labeled training set that supervised learning requires. GNNs enable direct analysis of smart contract bytecode without any manual feature engineering, predefined patterns, or expert rules.

SCooLS is the first application of semi-supervised learning to smart contract vulnerability analysis, as well as the first deep learning-based vulnerability analyzer to identify specific vulnerable functions. SCooLS's performance is better than existing tools, with an accuracy level of 98.4%, an F1 score of 90.5%, and an exceptionally low false positive rate of only 0.8%. Furthermore, SCooLS is fast, analyzing a typical function in 0.05 seconds.

We leverage SCooLS's ability to identify specific vulnerable functions to build an exploit generator, which was successful in stealing Ether from 76.9% of the true positives.

## I. Introduction

Bugs in smart contracts can cause staggering losses [38], [42]. Accordingly, the research community has developed many static and dynamic analysis techniques [12], [18], [21], [26], [30]–[32], [32], [44]–[47] to identify vulnerabilities in smart contracts. These tools are highly impressive, but they depend on expert-crafted rules and manually engineered features, which makes such tools challenging to maintain and update. Moreover, most techniques require (or at least benefit from) source code. Unfortunately, source code is only available for a minority of contracts [5], and empirical observations [26], [32] suggest that bytecode analysis is crucial.

SCooLS, our <u>S</u>mart <u>C</u>ontract <u>L</u>earning (<u>S</u>emi-supervised) engine uses deep learning rather than static/dynamic analysis to flag vulnerabilities. Previous applications of machine learning to this domain yielded promising results [5], [25], [29], [36], [40], [41], but this line of work is less developed than tools based on static or dynamic analysis. Only SoliAudit [25] and DLVA [5] have publicly-available tools.

SCooLS borrows two key ideas from DLVA: deep learning (neural nets) and, more specifically, the use of graph neural networks (GNNs). However, we differ in several critical ways: we use *semi-supervised learning* rather than *supervised learning*, and target *functions* rather than *contracts*. Our framework design and use of committees during training are also innovations. Accordingly, our resulting models are novel. Moreover, our focus on functions allowed the development of an *exploit generator*. SCooLS is the first tool to utilize these elements in a smart contract vulnerability analyzer.

SCooLS focuses on a single well-known smart contract vulnerability, "reentrancy-eth" (specifically, SWC-107 according to the <u>S</u>mart contract <u>W</u>eakness <u>C</u>lassification system [37]). We focus on reentrancy because it is a well-studied and serious vulnerability [27], [38]. This means that detecting reentrancy has value in practice; moreover, its popularity allows for a good comparison with previous work. Although most famous for the DAO attack in 2016, reentrancy continues to be a persistent issue in Ethereum smart contracts. Notable recent examples include the Uniswap/Lendf.Me hacks in April 2020, the SURGEBNB hack in August 2021, the CREAM FINANCE hack in August 2021, the Siren protocol hack in September 2021, and the Omni attack in July 2022 [9]. It took two weeks to find and hand-verify our core *ReentrancyBook* data set (described in §III). Extending SCooLS to other vulnerabilities would require similar incremental effort.

§II We provide background information.

§III **Contribution 0:** we assemble and hand-verify the *ReentrancyBook* data set, containing 22 vulnerable and 480 non-vulnerable functions. We also collect and preprocess our large unlabeled *BigBook* data set.

§IV We explain the design of our smart contract vulnerability analyzer SCooLS. **Contribution 1:** SCooLS is the first machine learning-based technique to identify specific vulnerable functions rather than labelling the contract as a whole. **Contribution 2:** we use semi-supervised learning to train SCooLS. Semi-supervised learning helps address the scarcity of high-confidence labeled code data in practical smart contract vulnerability classification tasks. In total we train 120 distinct models derived from applying a variety of hyperparameters to five state-of-the-art graph neural networks, using a voting system to smooth out the variance in individual models during training.

§V **Contribution 3:** we implement an exploit generator to prove that the detected vulnerabilities can be exploited by attackers to steal contract funds.

§VI We measure the performance of SCooLS and compare it with three state-of-the-art tools. SCooLS dominates the competition, obtaining a higher accuracy level of 98.4%,

a higher F1 score of 90.5%, and the lowest false positive rate of just 0.8%. Moreover, the analysis is fast, requiring only 0.05 seconds per function. We also showed that the exploit generator was able to attack 76.9% of the true positive instances for which an ABI was available.

§VII We discuss related work and conclude.

**SCooLS availability and ethical considerations.** Any vulnerability analyzer can be used with ill intent. Blockchains are tricky for responsible disclosure [7]. Attackers are incentivized to find and attack weak contracts, and due to the pseudonymous nature of the blockchain, it is hard to quietly inform participants of vulnerabilities. Concerningly, SCooLS allows attackers to target weak contracts relatively precisely, and our Exploit Generator enables nearly automatic theft.

On the other hand, reentrancy has long been studied and many available tools already flag it (*e.g.*, [5], [12], [31], [45]). Other exploit generators for smart contracts have also been published and made publicly available (*e.g.*, [22], [24]). Moreover, honest actors benefit from SCooLS too: everyone wants to know if the contracts they use are vulnerable.

To balance these considerations, we will release SCooLS *without the Exploit Generator* 60 days after publication and the Exploit Generator a further 60 days after that: publication at BCCA on 24-26 October, 2023; SCooLS release on 25 December (Christmas) 2023; and Exploit Generator release on 23 February 2024. Researchers who wish to benchmark with SCooLS before these dates should contact the authors for access. **Contribution 4:** SCooLS will be available for download from **https://bit.ly/SCooLS-Tool** (see "README.md").

## II. BACKGROUND

*a) Ethereum:* Ethereum is a decentralized open-source blockchain, introduced in 2014 and operational on July 30th, 2015 [49]. Ethereum allows anyone to deploy and interact with immutable decentralized applications. Ethereum smart contracts are self-executing agreements between parties that run on the blockchain. Smart contracts run a variety of decentralized finance apps and services [8]. Ethereum has significantly risen in popularity over time: average daily transactions increased from 10K in January 2016 to 1M+ in January 2023 [51]. Smart contracts have become more valuable as Ethereum has grown, making them attractive to attackers.

*b) Re-entrancy:* The semantics of the Etherum Virtual Machine (EVM) bytecode can be subtle. Developers may not understand that sending ether to a contract can trigger a function call; or that a non-recursive function can be re-entered before termination. In a reentrancy attack, a malicious actor takes advantage of these two subtleties to repeatedly call a vulnerable withdrawal-type function. If the contract has been naïvely coded, this can result in the same withdrawal being executed multiple times, despite the programmer's informal intention to authorize only one withdrawal. The quintessential Ethereum reentrancy attack occurred in 2016, when a hacker exploited a vulnerability in the smart contract code of the DAO to steal over 3.6 million Ether [27], [38].

*c) Deep learning styles:* The two most popular deep learning techniques are *supervised* and *unsupervised*. Supervised learning involves training a model on a *labeled* dataset, attempting to induce relationships between the elements and their labels. The main disadvantage of supervised learning is the need to source a large amount of labeled data for training, which can be expensive and time-consuming to collect. Moreover, labeling data can be subjective and error-prone, and mislabeled data can affect the accuracy of the trained model. For example, if we use a static analyzer to label a large training data set, training must cope with the false positives and negatives produced by said analyzer.

In contrast, unsupervised learning involves analyzing *unlabeled* data to identify patterns within the data, such as clusters, anomalies, or associations. Avoiding the necessity oracle-labeled data is a major positive, but without a target output for comparing predictions, it can be difficult to develop a model to solve a particular decision problem of interest.

*Semi-supervised* learning strikes a balance between supervised and unsupervised learning. An initial set of models is trained on a small *labeled* data set, and then used to label a large *unlabeled* data set. The resulting labels are sorted by confidence, and the high-confidence labels are then used to train the next generation of models. The process then iterates.

*Key idea:* In many domains, labeled data is scarce or expensive to obtain. Semi-supervised learning uses one to three orders of magnitude less data than supervised learning, and it is easier to gain confidence in the labels of a small data set.

Semi-supervised learning leverages the small labeled training data set to orient its models (as compared to unsupervised learning). The large unlabeled data set is used to increase generality and improve robustness to noise/outliers (as compared to supervised learning on a small data set). By leveraging both, the model can learn to recognize patterns and make accurate predictions on new data, despite having only a small amount of genuinely high-confidence labeled data available.

*d) Graph Neural Networks:* A graph neural network (GNN) is designed to perform inference on graph data structures. In other words, it is a neural network that can classify based on graph features. In a GNN, each node in the graph is represented by a feature vector containing information about the node and its neighbors. The GNN uses a series of graph convolutional layers based on *message passing* [15] to update each node's feature vectors by aggregating the information from neighboring nodes. This process is iterated to allow the network to learn more complex relationships between nodes (*e.g.*, after three graph convolutional layers, a node has information about the nodes three steps away from it).

Graph neural networks (GNNs) have been successfully applied in a wide range of domains across various learning settings. *Key idea:* GNNs are particularly effective when dealing with datasets that can be represented as graphs, where traditional machine learning algorithms may not be suitable.

*e) Evaluative Metrics:* We evaluate the quality of our models from several complementary perspectives. We use metrics Accuracy (ACC), F1-Score (F1), and False Positives

Rate (FPR). Accuracy is the number of correct predictions made by the model divided by the total number of predictions made. However, accuracy alone can be misleading if the dataset is imbalanced or if the cost of false positives and false negatives are not equal. An F-score (F1) is commonly used to benchmark deep learning for classification tasks, defined as the harmonic mean of *precision* and *recall*. Precision measures the proportion of true positives out of all predicted positives, while recall measures the proportion of true positives out of all actual positives. The False Positive Rate represents the proportion of actual negative instances that are incorrectly predicted as positive by the model. In other words, it measures the percentage of times that the model generates a false positive prediction out of all negative instances.

## III. DESIGN OF DATA SETS

We assemble two data sets: a small high-confidence manually-labeled *ReentrancyBook*, and a large unlabeled *BigBook*. **Contribution 0.** We publish both data sets [1], [3].

*a) ReentrancyBook:* We collected 932 smart contracts, containing a total of 11,587 functions, from prior work [2], [13], [14]. We then removed redundant functions to yield 502 distinct functions. All of the 932 contracts had source available, enabling manual labeling of these 502 functions as *vulnerable* or *non-vulnerable* for the *reentrancy* bug.

Labelling functions requires judgment calls, and our labelling did not always agree with previous work. Not every vulnerable function is actually exploitable [35], and some functions are exploitable only under unusual circumstances. For example, some functions can only be run by the contract's owner; or will only send Ether to specific hardcoded addresses; or can only work at specific times or block numbers; or some piece of contract state is updated before reentrancy, preventing exploitation. *We only label a function vulnerable if it is exploitable by the general public without such restrictions.* In total, we consider only 22 functions to be vulnerable, with the remaining 480 considered non-vulnerable. We dub the unique labelled functions the *ReentrancyBook* data set.

We use the *stratified sampling* of *scikit-learn* [34] to divide *ReentrancyBook* into halves, with one half being the training/validation set *ReentrancyStudyBook* (250 functions, of which 11 are vulnerable), and the other as the test set *ReentrancyTestBook* (252 functions, of which 11 are vulnerable).

*b) BigBook:* Semi-supervised learning requires both a small trusted core data set (*ReentrancyBook*) and large secondary data set. We downloaded the latter on Feb 24, 2023 from Google BigQuery [16], yielding 17,806,779 contracts containing 76,024,596 functions[1]. 99.3% of functions returned from BigQuery are duplicates; removing redundant functions left us with 554,111 distinct functions. We further removed the 446 functions already contained in *ReentrancyBook* to leave us with 553,665 distinct functions, which we dub the *BigBook* data set. The *BigBook* functions are unlabeled, and by construction *BigBook* and *ReentrancyBook* are disjoint.

[1]The query is available here: https://console.cloud.google.com/bigquery?sq=814627022739:e4a5075f5a7141078f0e170ced82ffa0

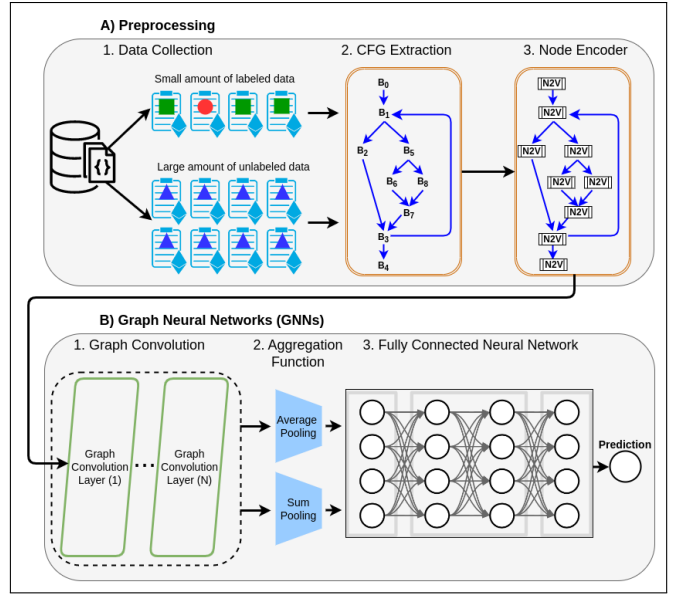

Fig. 1. Data Preprocessing and Graph Neural Networks (GNNs) Design.

## IV. DESIGNING SCOOLS

The design framework of SCooLS is sketched in Figure 1. The design divides into two parts: A) Preprocessing and B) Graph Neural Networks. Preprocesing is done at the beginning and once. The Graph Neural Networks (GNNs) are where the training occurs. We will discuss each part in turn.

### A. Preprocessing

Preprocessing begins with the data collection and manual labeling discussed above in §III, and then proceeds to turn a collection of contracts into a collection of vector-labeled graphs representing individual contract functions. **Contribution 1:** SCooLS is the first machine learning-based technique judging *individual functions* rather than *whole contracts*.

Conventional NLP pretraining techniques treat code as a sequence of tokens, just as they would a natural language. However, this approach neglects the valuable structural information present in code that can aid the understanding of its behaviour. Control-flow graphs (CFGs), directed graphs whose vertices are basic blocks and whose edges represent execution flow, are more useful for analysis because they capture important semantic structures within the contract [5].

We use *evm-cfg-builder* (v0.3.1) [33] to extract control flow graphs (CFGs) from directly EVM bytecode. The average function in our data sets has 14 basic blocks (nodes), each containing a textual sentence of opcodes (*e.g.*, "PUSH1 0x80 PUSH1 0x40 MSTORE CALLVALUE...").

Most machine learning techniques prefer to work on vectors rather than sentences, so to encode a node into fixed-length 512-dimensional vector (N2V), we use the transformer architecture [48] in the Universal Sentence Encoder [10].

Transformers rely on the self-attention mechanism, processing the whole sequence all at once (no sequential processing like in RNNs), and assigning a weight to each opcode to indicate how much "attention" the model should pay to said opcode. The model takes opcode order and the larger surrounding context into account when generating an opcode representation. The transformer encoder is composed of 6 stacked transformer layers. Each layer has two sub-layers: a multi-head self-attention mechanism followed by a fully connected feed-forward network. Transformer uses a residual connection around each of the two sub-layers, followed by layer normalization to produce its 512-dimensional outputs.

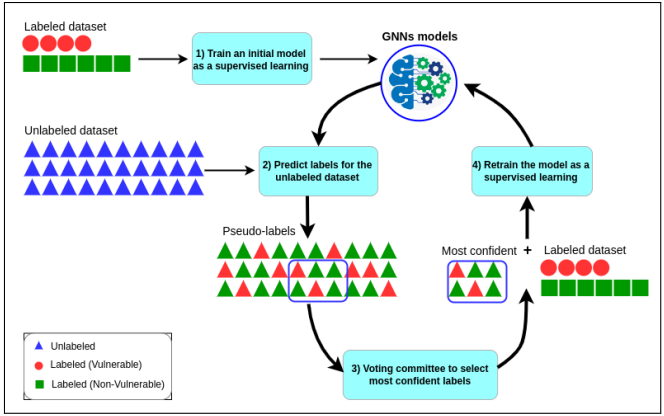

Fig. 2. The Smart Contracts Semi-Supervised Learning (SCooLS).

### B. Graph Neural Networks (GNNs)

SCooLS's neural nets are based on five state-of-the-art graph convolution methods [19], [23], [43], [50], [52]. As shown in Figure 1.B, our design consists of a series of graph convolutional layers, followed by an aggregation, which in turn is followed by a fully connected feed-forward network with a sigmoid activation function for classification. The design space of our models involves several hyperparameters, including:

1) the number of graph convolutional layers (1, 2, or 3)
2) the number of neurons (32, 64, 128, or 256) updated by each layer using a rectified linear unit (ReLU) activation.
3) the pooling aggregation function (average or sum).
4) the number of neurons in two Dense layers with a ReLU activation in the fully connected feed-forward network, which is the same as the number of neurons in the convolutional layer. Each Dense layer is surrounded by BatchNormalization and Dropout layers to enhance the model's generalization and prevent overfitting.

*Key idea:* our framework thus generates $5 \times 3 \times 4 \times 2 = 120$ models that can learn from different perspectives, resulting in a significant improvement in accuracy.

### C. Semi-Supervised Self-Training

With the basic design framework in place, we must now explain how the neural nets are trained. **Contribution 2:** We overview SCooLS's training cycle in Figure 2, illustrating the semi-supervised learning process. The process begins by using our small labeled dataset *ReentrancyStudyBook* to train an initial set of models (the training engineering is in §VI).

After the initial models are trained, we enter an iteration loop in which we ask the current set of models to judge the enormous *BigBook* of 553,665 functions. The output for each of the 120 models is a number between 0 (certainly non-vulnerable) and 1 (certainly vulnerable). *Key idea:* we now run a voting committee to determine which contract labels have sufficient model support. We discard any model with confidence $0.1 < x < 0.9$, leaving only the high confidence models, which then vote for 0 or 1. If a function gets 80+ votes (two-thirds of the models), then that label is accepted. A function without such a supermajority is considered *unknown*.

We take the newly-labeled functions, add the trusted *ReentrancyStudyBook*, and retrain the models for the next iteration. We continue until a termination condition is met.

### D. Final result

After training, analyzing a fresh contract is straightforward. The bytecode is converted to a CFG, whose nodes are transformed into vectors. Subsequently, the trained GNNs models from the final self-training iteration predict labels for each function, producing a total of 120 predictions. Finally, the voting committee applies a voting mechanism to determine the final classification label based on whether at least two-thirds of the predictions with a confidence level of 90% or higher classify the function as *vulnerable* or *non-vulnerable*.

## V. EXPLOIT GENERATOR DESIGN

Much of the effort in a hack is *finding* the needle vulnerability in the blockchain haystack; as we will see in §VI, SCooLS does an admirable job at this. Step two, *exploiting* the discovered vulnerability, is then often straightforward.

A *smart contract exploit generator* automatically generates the malicious code needed to exploit a vulnerable smart contract. **Contribution 3:** we implement a simple yet effective exploit generator to prove that attackers can exploit the detected reentrancy vulnerabilities to steal contract funds.

*a) Generating the exploit:* The generator is given the Application Binary Interface (ABI) of the victim contract $C$, together with the name of the function **V** that SCooLS has flagged as vulnerable. The generator examines the ABI to identify all functions $\mathbf{P}_1, \ldots, \mathbf{P}_n$ marked as `payable`. Next, for each $\mathbf{P}_i$, the generator proceeds as follows:

1) It generates an `Itarget` interface to facilitate interaction with the victim contract $C$ by combining the signature of the target payable function $\mathbf{P}_i$ with the signature of the flagged-vulnerable function **V**.
2) It generates the contract `TheAttacker`, containing the functions `attack_step1`, `attack_step2`, `receive`, and `steal`, together with contract boilerplate. The details are generated by template as follows.
3) In `attack_step1`, the attacker contract invokes the chosen `payable` function $\mathbf{P}_i$. Appropriate-typed arguments are selected at random from predefined dictionary values. The result is to transfer some Ether to the victim contract, emulating an honest user's deposit.

4) In `attack_step2`, the attacker contract calls the vulnerable function **V**, again providing it with suitably-typed and randomly-chosen arguments.
5) The key to the hack is the `receive` function, which is automatically triggered if **V** transfers Ether to `TheAttacker` during `attack_step2`. `receive` checks to see if $C$ has sufficient remaining funds to justify continuing the attack, and if so calls **V** again[2].
6) Lastly, the `steal` function transfers the money out of `TheAttacker` contract and into the hacker's wallet.

Here is the attack for victim contract `TheBank` using payable function `deposit` and vulnerable function `withdrawal`:

```solidity
pragma solidity >=0.1.0 <0.9.0;
interface Itarget{
  function deposit() external payable;
  function withdrawal() external; }
contract TheAttacker {
  Itarget public theBank;
  address payable public attacker;
  uint256 public amount = 1 ether;
  constructor(address _thebankAddress,
    address payable _attackerAddr) {
  theBank = Itarget(_thebankAddress);
  attacker = _attackerAddr; }
  function attack_step1() external payable {
    theBank.deposit{value: msg.value}(); }
  function attack_step2() external {
    theBank.withdrawal(); }
  receive() external payable {
    if (address(theBank).balance >= amount) {
      theBank.withdrawal(); } }
  function steal() public payable{
    attacker.transfer(address(this).balance); } }
```

*b) Testing the exploit:* We use Ganache [20], a local development and testing platform, to replicate the blockchain environment. For each generated attacker contract, we:

1) Deploy the bytecode and ABI of the victim contract to our local blockchain, with a balance of 0 Ether.
2) Allow normal users to perform standard transactions, invoking the *payable* function to send some Ether to the victim contract and increase its balance.
3) Deploy `TheAttacker` to the blockchain, and initiate the hack by executing the *attack_step1*, *attack_step2*, and *steal* functions in sequence using three transaction calls.

We compare the attacker's balance before and after the attack to see if the attack was successful:

```
Attacker Balance Before Attack :   1808.65   Ether(s)
Attacker Balance After Attack  :   1810.64   Ether(s)
This exploitation net profit is:   1.99      Ether(s)
The detected reentrancy can be exploited.
```

## VI. EXPERIMENTS AND EVALUATION

### A. Experimental setup

Our machine had 32 GB of memory and a 12-core 3.2 GHz Intel(R) Core(TM) i7-8700 CPU. We used 64-bit Ubuntu 20.04.6 LTS (Focal Fossa), tensorflow 2.12.0 [4], tensorflow_hub 0.13.0, spektral [17], evm-cfg-builder [33], and ganache-2.5.4-linux-x86_64.AppImage [20].

[2]A more sophisticated version would also track the recursion depth and would halt the theft before hitting the 1,024 function call depth.

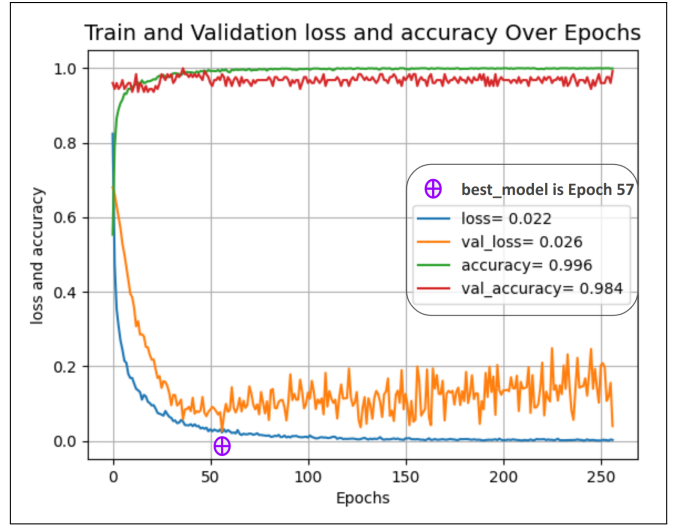

Fig. 3. Training Accuracy and Loss, and Validation Accuracy and Loss

We train the initial models using *ReentrancyStudyBook*, and then begin the self-training cycle with a learning rate of 0.005, a batch size of 2,048, and 1,000 epochs. We use the Adam optimizer with a categorical cross-entropy loss function.

To prevent overfitting, we use a rolling 200-epoch window and measure the validation loss for each model against the *ReentrancyStudyBook*. If the model with the lowest validation loss was 200 epochs ago, then training stops and restores that model. Figure 3 displays the training and validation accuracy of an arbitrarily-chosen model. The validation loss (orange line) minimizes around epoch 57, so when training stops 200 epochs later, it restores the model from epoch 57.

TABLE I
THE *BigBook* DATA SET.

| Models | Vulnerable | Non-vulnerable | Unknown |
|---|---|---|---|
| *BigBook* size | 0 | 0 | 553,665 |
| *ReentrancyStudyBook* training | 48 | 529,783 | 23,834 |
| Self-training first iteration | 69 | 553,132 | 464 |
| Self-training second iteration | 82 | 553,322 | 261 |

After training 120 models (one cycle depicted in Figure 2), we repeat again, for three iterations in total including the initial training. Table I shows how the training process shrinks the unknown set over time. In total we trained 360 models over a period of four days, *i.e.* approximately 16 minutes per model. The 120 models from the last iteration are used in SCooLS.

### B. SCooLS vs. state-of-the-art tools

To test the performance of our approach, we evaluated it on the *ReentrancyTestBook*, which by construction is disjoint from the *ReentrancyStudyBook* and *BigBook* data sets used during training. Moreover, we manually labelled each of its 252 functions, giving us high confidence their labels.

We also wish to compare SCooLS against the state-of-the-art actively-used alternatives ConFuzzius v0.0.1 [45], Slither 0.9.0 [12], and Mythril v0.23.17 [31].

The results are shown in Table II. The primary data are True Positives (TP), False Positives (FP), True Negatives (TN), and False Negatives (FN). The derived statistics are Accuracy (ACC), F-score (F1), and False Positive Rate (FPR).

SCooLS has an overall accuracy 98.4% and F1 score 90.4% with an associated false positive rate of only 0.8%. It enjoys the highest accuracy among the tools, the highest F1 score, and the lowest false positive rate. Moreover, the average time to analyze a function was only 0.05 seconds, tied for first place.

ConFuzzius [45] uses a hybrid fuzzer that uses data dependency analysis to generate effective test cases for smart contracts. Hybrid fuzzing involves an initial stage of traditional fuzzing that continues until reaching a saturation point where no new code coverage is achieved after executing a predefined number of steps. Upon reaching this point, the hybrid fuzzer automatically switches to the process of symbolic execution, which performs an exhaustive search for unexplored branching conditions. If a branching condition is found, the symbolic execution process solves it, and the hybrid fuzzer reverts to the traditional fuzzing stage. ConFuzzius attains an accuracy level of 97.2%, accompanied by an F1 score of 85.3%, while maintaining a remarkably low false positive rate of merely 2.1%. The entire process takes 0.48 seconds per function.

Slither [12] is a static analysis framework that analyzes smart contracts source code through the integration of data flow analysis and taint analysis. Slither incorporates a large number of detectors that can identify specific issues such as reentrancy, integer overflow, and uninitialized variables. Slither achieves an accuracy level of 95.6%, accompanied by an F1 score of 82.2%, while retaining a notably low false positive rate of 4.6%. The complete analysis process for each function takes a mere 0.05 seconds, also tied for first place.

Mythril [31] is a powerful tool for identifying potential security issues in smart contracts, and it is widely used by developers, auditors, and researchers. Mythril works by using symbolic execution to explore all possible execution paths through the smart contract's code to detect potential vulnerabilities. Mythril demonstrates an accuracy level of 92.1%, along with an F1 score of 72.8%, while maintaining a relatively higher false positive rate of 7.9%. The complete analysis process for each function is relatively slower, taking around 12.5 seconds per function.

TABLE II
SCooLS VS. STATE-OF-THE-ART TOOLS.

| Tool | TP | FP | TN | FN | ACC | F1 | FPR | Time |
|---|---|---|---|---|---|---|---|---|
| **SCooLS** | 9 | 2 | 239 | 2 | 98.4 | 90.5 | 0.8 | 0.05 |
| **ConFuzzius [45]** | 9 | 5 | 236 | 2 | 97.2 | 85.3 | 2.1 | 0.48 |
| **Slither [12]** | 11 | 11 | 230 | 0 | 95.6 | 82.2 | 4.6 | 0.05 |
| **Mythril [31]** | 10 | 19 | 222 | 1 | 92.1 | 72.8 | 7.9 | 12.5 |

### C. Exploit generator results

To test our exploit generator, we took the 82 positives SCooLS found in *BigBook* (see Table I). These 82 positives have a further 89 duplicate instances in our Google BigQuery, for a total of 171 vulnerable functions.

Unfortunately, not all of the contracts containing these functions offered an Application Binary Interface (ABI) that allows the general public to easily interact with their functions. Specifically, only 33 out of the 171 functions had an ABI available on Etherscan, yielding 14 unique functions. We were able to retrieve the source code for all 14 unique functions. Manual inspection resolved 4 false positives (7 including duplicates, 21.2%) and 10 true positives (26 including duplicates, 78.8%).

Most of the false positives are due to functions that are *nearly* exploitable, for example because:

1) the contract cannot receive ether, so nothing to steal;
2) the receiver is hardcoded into the contract, so ownership of a specific address is required to attack; or,
3) the function can only be called at certain timestamps or block numbers, which makes it challenging to exploit.

Very few of the flagged functions are not exploitable for the "right" reasons, *i.e.* because the contract manages its internal state shrewdly to avoid the exploit during reentrancy.

Of course, our exploit generator cannot exploit a false positive. However, it was able to exploit 6 of the true positives (20 including duplicates, 76.9%) as shown in Table III. The remaining 4 true positives (6 with duplicates) are exploitable, but our exploit generator is not smart enough to do so.

TABLE III
THE EXPLOIT GENERATOR.

| Address | Vulnerable function | Function duplicates | Exploit Gen |
|---|---|---|---|
| 0x65e5909d665cb ... | CashOut(uint256) | 11 | ✓ |
| 0xe610af01f92f1 ... | Collect(uint256) | 1 | ✗ |
| 0x2ec17d1df257d ... | call() | 1 | ✗ |
| 0x2a98d8fc14b31 ... | withdraw() | 2 | ✓ |
| 0xa5d6accc56953 ... | CashOut(uint256) | 4 | ✓ |
| 0x0ebe1a9cbf4e2 ... | settleEther() | 2 | ✗ |
| 0xdd17afae8a3dd ... | Collect(uint256) | 2 | ✗ |
| 0xb7c5c5aa4d429 ... | withdraw() | 1 | ✓ |
| 0xf6dbe88ba55f1 ... | withdraw(uint256) | 1 | ✓ |
| 0xaf905ab8dad7c ... | pullFunds() | 1 | ✓ |

## VII. RELATED WORK, FUTURE WORK, AND CONCLUSION

### A. Related work

*a) Static and dynamic analyzers:* Detection of software vulnerabilities ensures their security and reliability. Traditional methods, such as static and dynamic analysis, have long been applied to smart contracts [12], [18], [21], [26], [30]–[32], [32], [44]–[47]. Several of these tools are under active development and are widely used in the community. These tools use expert-crafted rules and manually engineered features, which can make it challenging to maintain and update them. Many (although not all) require or at least benefit from source code. Bytecode analyzers tend to be less precise (and are often significantly slower), so there is a need for more advanced techniques for vulnerability detection at scale.

*b) Machine learning:* Machine learning approaches for smart contract vulnerability detection has gained some attenion as an alternative to traditional analyzers [5], [25], [29], [36], [40], [41]. Wesley *et al.* [41] improved vulnerability detection in smart contracts by using a customized LSTM neural network that sequentially examined opcodes, resulting in superior accuracy compared to Maian. Momeni *et al.* [29] proposed a machine learning model that used AST and CFG to analyze static source code, with 17 code complexity-based features for model training. This approach achieved a faster processing time and lower miss rate than the static analyzers Mythril and Slither. Liao *et al.* [25] created SoliAudit, an approach that combines machine learning and a dynamic fuzzer to enhance vulnerability detection capabilities. To construct the feature matrix, Liao *et al.* used word2vec [28] to generate vectors for each opcode, which were then combined row-wise; they did not take control flow into account. Sun *et al.* [40] improved smart contract vulnerability detection by incorporating an attention mechanism into (non-graph) convolutional neural networks, outperforming Oyente and Mythril in both miss rate and processing time. The proposed method utilizes a CNN model combined with self-attention and achieves swift detection of vulnerabilities with a reduced miss rate and average processing time. Qian *et al.* [36] proposed a deep machine learning approach to identify Reentrancy vulnerabilities in smart contracts, utilizing the BLSTM-ATT model. Their method divided the source code into snippets and employed word2vec to extract code features, showing that deep learning techniques are suitable for smart contract vulnerability detection and can achieve high performance.

DLVA [5], [6], a deep learning-based tool designed to analyze vulnerabilities in smart contract bytecode, is the closest to SCooLS, since we build on their use of GNNs and insight that we can train a bytecode analyzer using labeled source code. DLVA employs a comprehensive approach that can detect 29 distinct vulnerabilities in smart contracts, rather than the single vulnerability we focus on. In contrast to SCooLS, DLVA uses supervised learning rather than the semi-supervised learning. Training labels are generated by Slither, which means that its model must cope with false positives and negatives in the training set. In addition, DLVA builds a single model, rather than a committee of 120 models as we do. DLVA focuses on whole-contract classification, which allows it to detect vulnerabilities across function boundaries. On the other hand, it is unable to specifically locate the vulnerabilities within the contract, whereas we can pinpoint specific vulnerable functions. DLVA does not include an exploit generator.

Sun *et al.* [39] proposed the ASSBert framework, which utilizes BERT [11] for smart contract vulnerability classification to train from a small number of labeled Solidity source files. The framework uses active and semi-supervised learning approaches to improve the model's performance. ASSBert outperformed baseline methods such as Bert, Bert-AL, and Bert-SSL. However, ASSBert has not been compared with state-of-the-art tools and is not publicly available.

The application of deep learning approaches to smart contract vulnerability detection shows promise, and semi-supervised learning techniques may improve the accuracy of the models while reducing the cost of obtaining labeled data.

*c) Exploit generators:* teEther by Krupp *et al.* [24] generates exploits for suicidal and call injection vulnerabilities on the Ethereum platform by analyzing the binary bytecode of a smart contract. However, it is not designed to identify or exploit other common vulnerabilities, such as reentrancy.

Jin *et al.* [22] developed EXGEN, a tool that generates attack contracts with multiple transactions and tests their exploitability on a private blockchain using public blockchain values. It is publicly available but benchmarking is challenging due to its complex environment setup.

## B. Future work and conclusion

In this research paper, we have proposed a deep learning approach for Smart Contracts Semi-Supervised Learning (SCooLS) to detect vulnerable functions in Ethereum smart contracts at the bytecode level. Our approach incorporates semi-supervised learning and deep graph neural networks (GNNs) to analyze smart contract bytecode without any manual feature engineering, predefined patterns, or expert rules.

SCooLS outperforms existing state-of-the-art tools, achieving an accuracy level of 98.4% and an F1 score of 90.5%, while exhibiting an exceptionally low false positive rate of only 0.8%. Additionally, the analysis process for each function is also quicker than existing tools, requiring only 0.05 seconds.

We introduced a voting committee to ensure the integrity of newly labeled data during the self-training process and to avoid the spread of errors to subsequent iterations. We implemented an exploit generator to verify that the detected vulnerabilities can be exploited by attackers to steal contract funds.

To the best of our knowledge, our work is the first to propose a semi-supervised self-training method for detecting vulnerabilities in smart contracts bytecode at the function level. Our results demonstrate the effectiveness of the proposed approach and its potential to enhance the security of smart contracts. In conclusion, our work contributes significantly to the field of smart contract security and provides a strong foundation for future research in this area.

We wish to explore several directions in the future. We aim to extend with training more vulnerabilities as mach as we can get a small labeled functions for them. We hope to enhance the Exploit Generator by using Fuzzing to help create function inputs for functions, possibly with a ML model to guide the associated search and make it more effective.

**Acknowledgements.** We thank Joxan Jaffar for his support.